\documentclass[pss]{wiley2sp}
\usepackage{amsmath}
\usepackage{bm}
\usepackage{w-greek}
 \usepackage[]{graphicx}
\usepackage{epsfig}

\begin{document}
\title{Entanglement of flux qubits through a joint detection of photons}

\titlerunning{Entanglement of flux qubits through a joint detection of photons}

\author{%
Marcin Kurpas\textsuperscript{\textsf{\bfseries \Ast}}
and El\.zbieta Zipper}

\authorrunning{M. Kurpas, E. Zipper}

\mail{e-mail
    \textsf{mkurpas@us.edu.pl}, Phone: +49 30 470 31 331, Fax: +49 30 470 31 334
    }

\institute{%
    Institute of Physics, University of Silesia, 40-007 Katowice , Poland}

\received{XXXX, revised XXXX, accepted XXXX}
\published{XXXX}


\abstract{%
\abstcol{%
We study the entanglement creation between two flux qubits interacting with electromagnetic field modes. No direct interaction between the qubits exists. Entanglement is reached using entanglement swapping method by an interference measurement performed on photons. } 
{ We discuss the influence of off-resonance and multi-photon initial states on the qubit-qubit entanglement. The presented scheme is able to drive an initially separable state of two qubits into an highly entangled state suitable for quantum information processing.}}

\titlefigurecaption{Selected covers of pss (a), (b), and (c) as an example for the \emph{optional} abstract figure (with or without caption). If there is no figure here, the text should fill both columns.}

\maketitle
\section{Introduction}
Recently, quantum information and computation are one of the fastest expanding areas of modern physics. The basic element of a quantum computer is a two state system, usually referred to us as a qubit. Several two-states systems has been proposed as candidates to build the qubit (atoms, ions \cite{har}, photons and solid state systems \cite{nakamura,orlando,migliore}). However, the power of quantum computing lies in coherent, parallel operations performed on many qubits that can be manipulated, coupled and read out in a controllable and non-destructive way. The second determinant is the scalability of the system.  As far the best candidates fulfilling these conditions seem to be solid state qubits built on charge \cite{nakamura} and flux \cite{orlando,migliore} degrees of freedom.
Quantum entanglement of composite systems plays a major role in information processing. In the last few years great progress has been made in creation and manipulation of entangled states in systems based on solid state qubits \cite{blais,majer,kok}.	
In recent papers we have investigated the entanglement of remotely located superconducting qubits placed in quantum cavities \cite{zipper,kurpas} making use of entanglement swapping scheme \cite{zukow}. To show the idea we have calculated the degree of entanglement for the strict resonance between the qubits and the cavities and for two specific initial states.

In this paper we want to give a deeper insight into this phenomenon and study the coherent coupling of qubits for different, more general, physical situations. We perform a model investigations on recently proposed non-superconducting flux qubit \cite{zipper1} built on semiconducting quantum ring with a barrier. Persistent currents running in opposite directions form the basis states and, when coupled by tunnelling, can lead to formation of a qubit.
We investigate entanglement for a set of initial states and for different resonance conditions. We show that entanglement of remotely located qubits through a joint detection photons is robust against detuning.
\section{The system}
The system consists of two non-superconducting, not directly interacting flux qubits $Q^1$ and $Q^2$ each placed in its own high quality quantum cavity $C^1$ and $C^2$ respectively. 
The quantized electromagnetic field enclosed inside the cavities interacts with the qubits. In the limit of the strong coupling regime ($g \gg \gamma_Q, \gamma_C$ \cite{haroche}, where $g$ is qubit-field coupling constant, $\gamma_Q$ ($\gamma_C$) is the qubit (cavity) decoherence rate) this interaction can be treated as coherent. For solid-state qubit-cavity systems this regime has been achieved in many experiments \cite{blais,majer}.

The Hamiltonian of the qubit-cavity ($QC$) system contains the qubit part $H_Q$, the cavity part $H_C$ and the interaction part $H_{int}$. In the basis in which the qubit states are diagonal the Hamiltonian can be written as
\begin{eqnarray} \label{h}
&H^k =& H^k_Q + H^k_C + H^k_{int}, \\
&H^k_Q =&\hbar \Omega_k \sigma_z, 
\end{eqnarray}
\begin{eqnarray} 
&H^k_C=&\hbar \omega_k\left( a^{\dagger }a+\frac 12\right), \\
&H^k_{int} =&\hbar g_k \left(a + a^{\dagger} \right) \sigma_x , \label{Hint}
\end{eqnarray}
where $ \Omega_k$ is the qubit frequency, $\omega_k$ is the cavity frequency, $g_k=I_0\sqrt{\frac{\hbar \omega_k L}2}$ is the coupling constant between the qubit and the electromagnetic field, $I_0$ is the current amplitude of the qubit , $L$ is the qubit inductance; $\sigma_i$ are Pauli matrices, $k=1,2$ numbers the qubit-cavity subsystems.\\
It is known \cite{geller} that flux qubits are strongly coupled to electromagnetic field and their interaction cannot be described by the Jaynes-Cummings ($JC$) model. Our estimations give $g=0.2$ (in units of $\omega$) for reasonable parameters.

We start from a separable initial state of $k$-th $QC$ subsystem 
\begin{equation}
\vert \psi(0) \rangle^k = \sum_{s=\downarrow,\uparrow} \gamma_{s} \vert s \rangle^k \otimes
\sum^\infty _{n=0} \beta_n \vert n \rangle^k  , 
\end{equation}
where $s=\left\lbrace \downarrow, \uparrow\right\rbrace $ denotes the qubit pseudo-spin states ($\uparrow$-ground ,$\downarrow$-excited), $\vert n\rangle$ are the photon number eigenstates, forming the so called Fock
 basis, $n=0,1,2,...$. In general, the dimension of photon space is infinite. Here we truncated the photon space to $N=10$ with the density matrix trace being at least 0.99.
The time evolution of $\vert \psi(0) \rangle^k$ in the absence of dissipation processes is generated by (\ref{h})
\begin{equation}
\vert \psi(t)\rangle^k = e^{(-i/\hbar H^k t)} \vert\psi(0) \rangle^k.
\end{equation}
The interaction between the qubit and the field couples their states and, in general, leads to entanglement creation
\begin{eqnarray}
|\psi(t)\rangle^1 &=&\sum_{n=0}^{9}[a_n(t)|\uparrow n\rangle^1+b_n(t)|\downarrow n\rangle^1 ], \\
|\psi(t)\rangle^2 &=&\sum_{n=0}^{9}[\tilde{a}_n(t)|\uparrow n\rangle^2+\tilde{b}_n(t)|\downarrow n\rangle^2 ].
\end{eqnarray}
As the two $QC$s subsystems do not interact with each other their time evolution goes independently and is described by 
\begin{equation}
\vert\psi(t)\rangle^{tot} = \vert \psi(t) \rangle^1 \otimes \vert \psi(t)\rangle^2.
\end{equation}

At this point we have two qubit-cavity subsystems (each in an entangled state: the qubit entangled with the electromagnetic field modes) being an analogue for two pairs of entangled photons in the original entanglement swapping experiment \cite{zukow}.
Then, to entangle two qubits one needs to entangle photons leaving the cavities by carrying out the Bell state measurement (BSM) at a certain moment $t'$. Quantum mechanically the BSM is described by the projection of $\vert \psi(t')\rangle^{tot}$ onto one of the Bell states e.g. onto the state $\vert \psi^{-}\rangle = 1/\sqrt{2}\left( \vert 01\rangle - \vert 10\rangle \right)$
\begin{equation}
\vert \psi(t')\rangle^{bsm} = \vert \psi^{-}\rangle\langle\psi^{-}\vert \psi(t')\rangle^{tot}.
\end{equation}
$\vert \psi(t')\rangle^{bsm}$ can be rewritten as
\begin{equation}
\vert \psi(t')\rangle^{bsm}= \frac{1}{\sqrt{2}}\vert \psi^-\rangle \otimes \vert \psi_{QQ}(t')\rangle,
\end{equation}
where $\vert \psi_{QQ}(t')\rangle$ is the qubit-qubit state vector after the measurement.

Our model calculations assume that the photons leave the cavities at the same time and their paths are totally indistinguishable. In the real entanglement swapping experiments this condition is difficult to be fulfilled. The partial distinguishability of the photons paths decreases the interference measurement contrast and as a result the experiment efficiency. However, there exists a narrow time window in which the photons can leave the cavities to keep their paths indistinguishable \cite{zukow,moehring}.\\
The unnormalized qubit-qubit state has the form 
\begin{equation}
\begin{array}{rcl}\label{psiQQ}
\vert \psi _{QQ}(t')\rangle & = &[a_0(t')\tilde{a}_1(t')-a_1(t)\tilde{a}_0(t')]|\uparrow \uparrow \rangle +  
\\
\ &\ & [a_0(t')\tilde{b}_1(t')-a_1(t')\tilde{b}_0(t')]| \uparrow \downarrow \rangle +
 \\
\ &\ & [b_0(t')\tilde{a}_1(t')-b_1(t')\tilde{a}_0(t')]| \downarrow \uparrow \rangle +
 \\
\ &\ & [b_0(t')\tilde{b}_1(t')-b_1(t')\tilde{b}_0(t')]| \downarrow \downarrow \rangle.
\end{array}
\end{equation}

The entanglement of this state can be quantified using e.g. concurrence proposed by Wooters \cite{wooters} which for pure states is defined as 
\begin{equation}
\tilde{C} = \mid\sum_i \alpha_i^2\mid,
\end{equation}
where $\vert \psi _{QQ}(t')\rangle = \sum_i \alpha_i \vert e_i\rangle$,
\begin{equation}
\begin{array}{rcl}
e_1 &=& 1/2\left(\vert\downarrow\downarrow \rangle -\vert \uparrow\uparrow \rangle\right), 
e_2=i/2\left(\vert \uparrow\uparrow\rangle -\vert \downarrow\downarrow\rangle\right),  \\
e_3 &=&i/2 \left(\vert \downarrow\uparrow\rangle + \vert \uparrow\downarrow\rangle\right) , 
e_4 = 1/2 \left(\vert\downarrow\uparrow \rangle - \vert \uparrow\downarrow\rangle\right).  
\end{array}
\end{equation}
This measure is an entanglement monotone and changes from 0 for disentangled to 1 for maximally entangled states.
\section{Results}
In our earlier papers (\cite{zipper,kurpas}) we discussed the entanglement for two specific initial states with single excitation. Here we want to test the swapping method for a range of initial states.

\begin{figure} [t]
\includegraphics*[width = 0.9\linewidth]{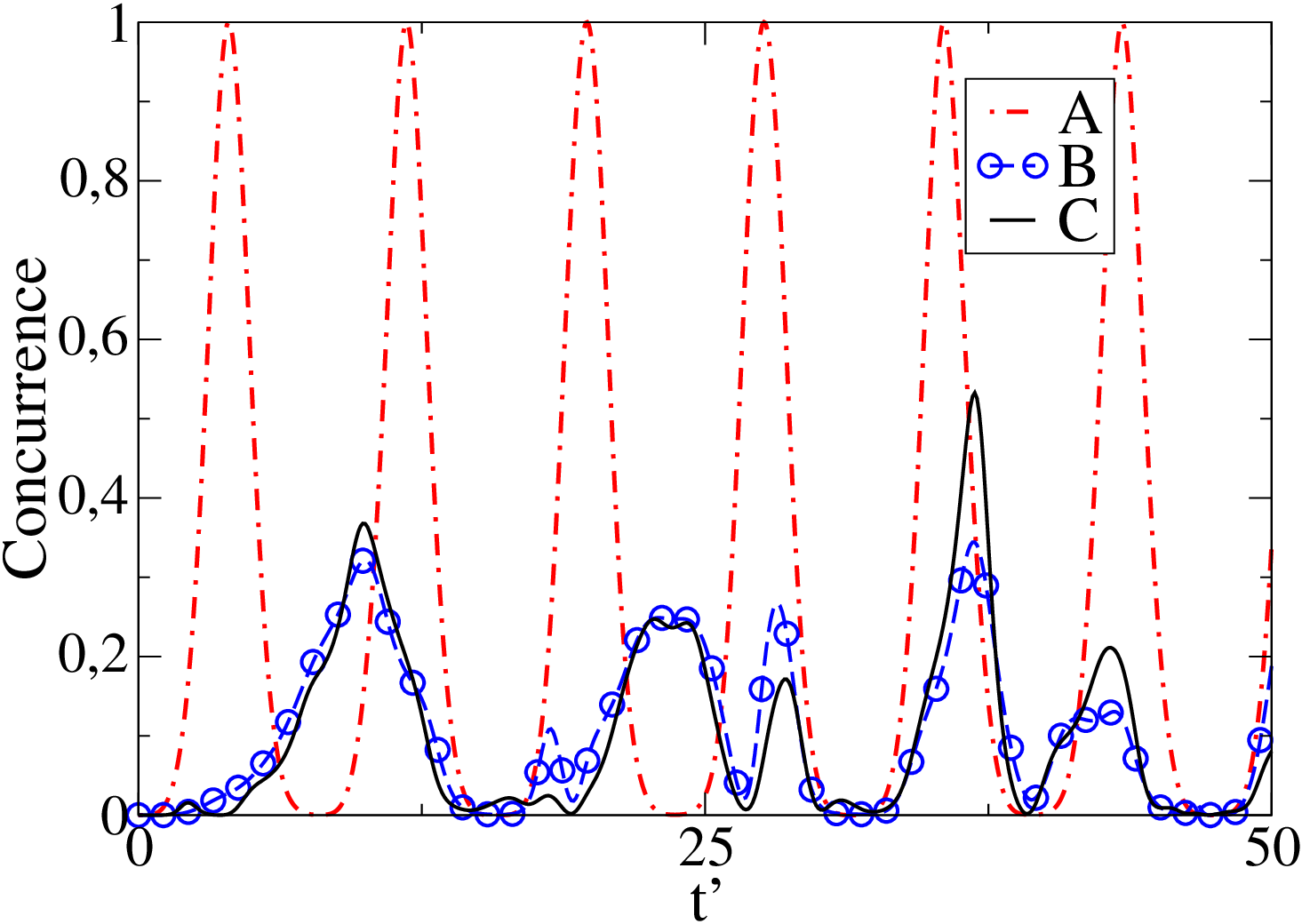}
\caption{(color online) The qubit-qubit (QQ) concurrence for different initial states:A - (\ref{e0g1}), B - (\ref{e01g01}), C - (\ref{e0123g0123}). The presence of additional photons in the cavity decreases the efficiency of entanglement swapping; $\omega_k = \Omega_k = 1$,$~g_k=0.2$.}
\label{f1}
\end{figure}

\begin{figure}[t]
\includegraphics*[width =0.9\linewidth]{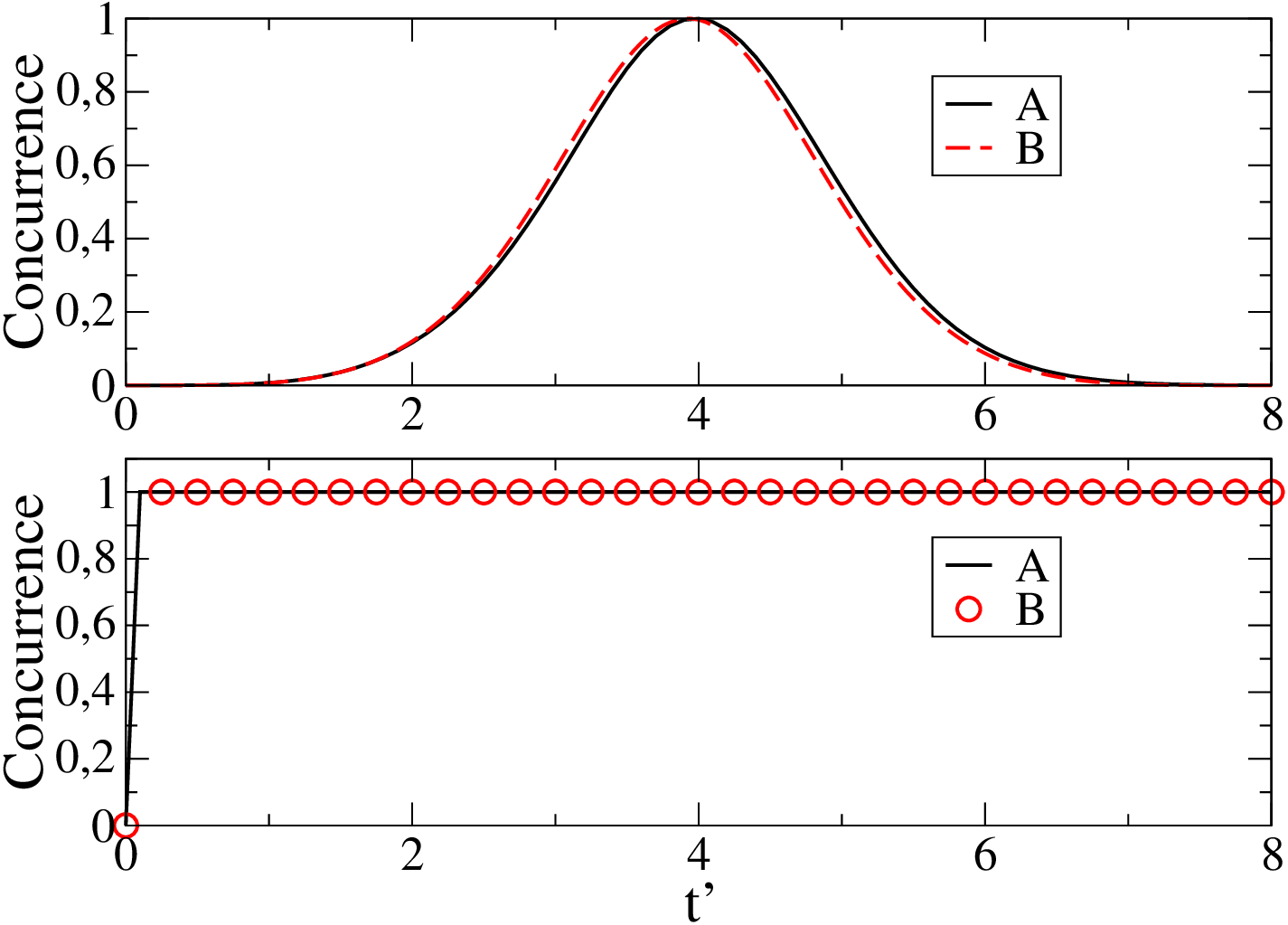}
\caption{(color online) The QQ concurrence for the JC (line A) and the Hamiltonian (\ref{Hint}) (line B). TOP: the initial state $\psi(0)\rangle^{tot}=\vert \downarrow 0 \rangle^1\otimes\vert \uparrow 1\rangle^2$. BOTTOM: the initial state $\psi(0)\rangle^{tot}=\vert \downarrow 0 \rangle^1\otimes\vert \downarrow 0 \rangle^2$; $\omega_k = \Omega_k = 1,~g_k=0.2$.}
\label{f2}
\end{figure}

At first we discuss the phenomenon when $(QC)^1$ and $(QC)^2$ subsystems start from different initial states (\ref{e0g1}) - (\ref{e0123g0123})
\begin{eqnarray}
\vert \psi(0)\rangle^{tot}&=&\vert \downarrow 0\rangle^1 \otimes\vert\uparrow 1\rangle^2,  \label{e0g1} \\
\vert\psi(0)\rangle^{tot} &=& \frac{1}{2}\left( \vert \downarrow 0\rangle^1 +\vert \downarrow 1\rangle^1 \right) \otimes
\left( \vert \uparrow 0\rangle^2 +\vert \uparrow 1\rangle^2 \right),  \label{e01g01}\\
\vert \psi(0)\rangle^{tot} &=& \frac{1}{4} \left( \sum_{n=0}^3 \vert \downarrow n\rangle^1 \otimes  \sum_{m=0}^3 \vert \uparrow m \rangle^2 \right).  \label{e0123g0123}
\end{eqnarray}
The interaction term $H_{int}$ establishes selection rules for the qubit-field dynamics
\begin{eqnarray}
\vert \downarrow m\rangle \rightarrow \vert \uparrow n\rangle, ~~\vert \downarrow n\rangle \rightarrow\vert\uparrow m\rangle, \\
\vert \uparrow m\rangle \rightarrow \vert \downarrow n\rangle, ~~\vert \uparrow n\rangle \rightarrow\vert\downarrow m\rangle, 
\end{eqnarray}
where $m$ and $n$ are even and odd numbers respectively.
This causes a strong dependence of the resulting entanglement on the initial state. For the initial state (\ref{e0g1}) (line A in Fig.\ref{f1}) the concurrence oscillates between the minimal and  the maximal value and has a periodic character. 
Because of the selection rules the system can evolve only to the $\vert \psi(t)\rangle^{tot} = \sum_{m,n}\left( \alpha_m\vert \downarrow m \rangle + \alpha_n \vert \uparrow n\rangle \right) $ ($m$-even, $n$-odd numbers) states. Those with numbers of photons larger than 1 are cut off at the BSM and only the states $\vert \downarrow 0\rangle$ and $\vert \uparrow 1\rangle$ give rise to (\ref{psiQQ}). It is also the reason why this result does not differ substantially from the result obtained for the $JC$ model (see top pane of Fig.\ref{f2}), even if the value of $g$ is ten times higher than the $JC$ limit \cite{geller}. 

For more general initial states (\ref{e01g01}),(\ref{e0123g0123}) although the selection rules are still present, the information about the $QC$ system is encoded in larger number of states. Most of them are cut off at the level of the BSM, resulting in decrease of entanglement of the qubits (curves B and C in Fig.\ref{f1}).


At second we study the entanglement for the case when  $(QC)^1$ and $(QC)^2$ start from the same initial states given by (\ref{e0e0}) - (\ref{e0123e0123})
\begin{eqnarray}
\vert \psi(0)\rangle^{tot}&=&\vert \downarrow 0\rangle^1 \otimes\vert\downarrow 0\rangle^2 \label{e0e0},  \\
\vert\psi(0)\rangle^{tot} &=& \frac{1}{2}\left( \vert \downarrow 0\rangle^1 +\vert \downarrow 1\rangle^1 \right) \otimes
\left( \vert \downarrow 0\rangle^2 +\vert \downarrow 1\rangle^2 \right), \label{e01e01}\\
\vert \psi(0)\rangle^{tot} &=& \frac{1}{4} \left( \sum_{n=0}^3 \vert \downarrow n\rangle^1 \otimes  \sum_{m=0}^3 \vert \downarrow m \rangle^2 \right). \label{e0123e0123}
\end{eqnarray}
In these cases $\vert \psi(t)\rangle^1 = \vert \psi(t)\rangle^2$ and the qubits get maximally entangled ($\tilde{C}$=1, bottom pane of Fig.\ref{f2}B) for almost every $t'$ (with some exceptions discussed in detail in \cite{zipper,kurpas}). The results are the same for all three initial states, even more, they fit the results for the Jaynes-Cummings model. This is the case only for identical and non-dissipative $QC$s what, in fact, is possible only in theoretical considerations. However we show below that strong entanglement can also be obtained for less restrictive conditions. 

\begin{figure}[t]
\includegraphics*[width = 0.9\linewidth]{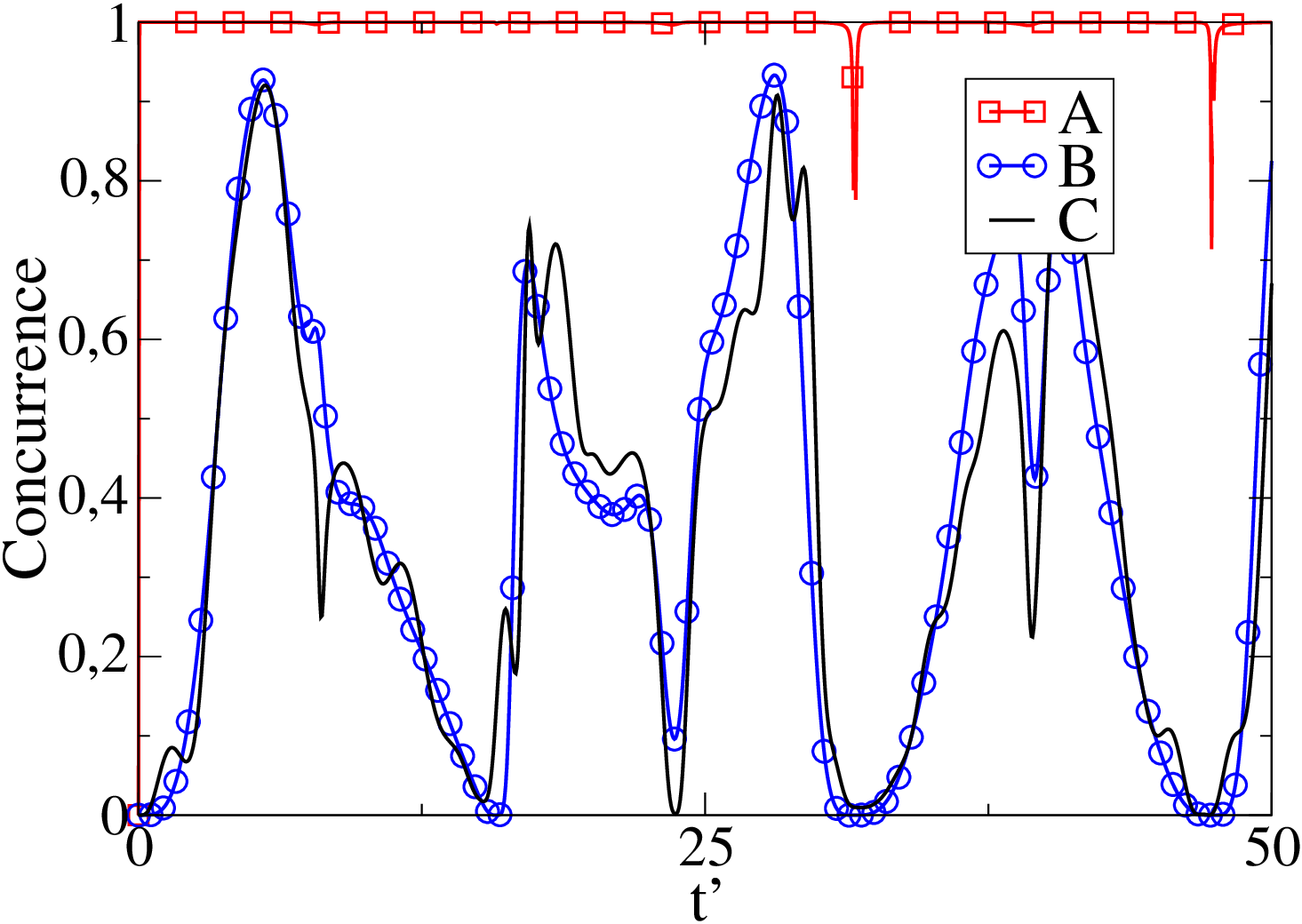}
\caption{(color online) The QQ concurrence for different initial states: A - (\ref{e0e0}), B - (\ref{e01e01}), C - (\ref{e0123e0123}). The presence of additional photons in the cavity decreases the efficiency of entanglement swapping; $\omega_1 = \Omega_1 = 1,\omega_2 = \Omega_2 = 0.95 $,$~g_k=0.2$.}
\label{f3}
\end{figure}

In spite of increasing possibilities of material engineering the problem of getting two identical samples is still the fact. To be more realistic now we take the qubits with different frequencies. We assume, that each cavity has the frequency equal to the qubit frequency $\Omega_k = \omega_k$. We see from Fig.\ref{f3} that for the initial states (\ref{e0e0}) - (\ref{e0123e0123}) we still get strong entanglement.

\begin{figure}[t]
\includegraphics*[width = 0.9\linewidth]{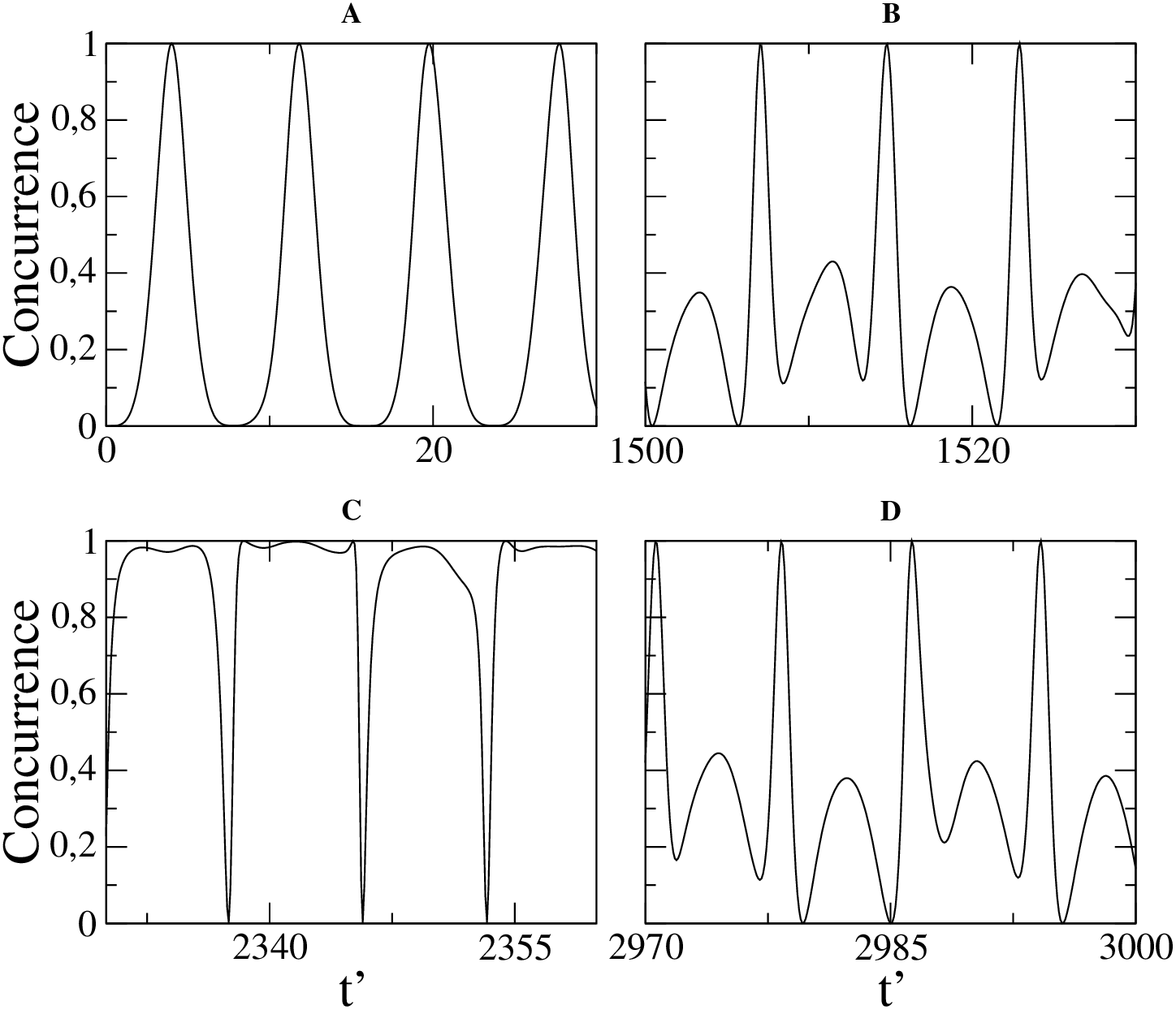}
\caption{The result of entanglement swapping for different values of the parameters $\omega_2 = \Omega_2 = 0.8$, $\omega_1=\Omega_1 = 1,~g_k=0.2$, The initial state is $ \psi(0)\rangle^{tot}=\vert \downarrow 0 \rangle^1\otimes\vert \uparrow 1\rangle^2$.}
\label{f4}
\end{figure}

\begin{figure}
\includegraphics*[width =0.9\linewidth]{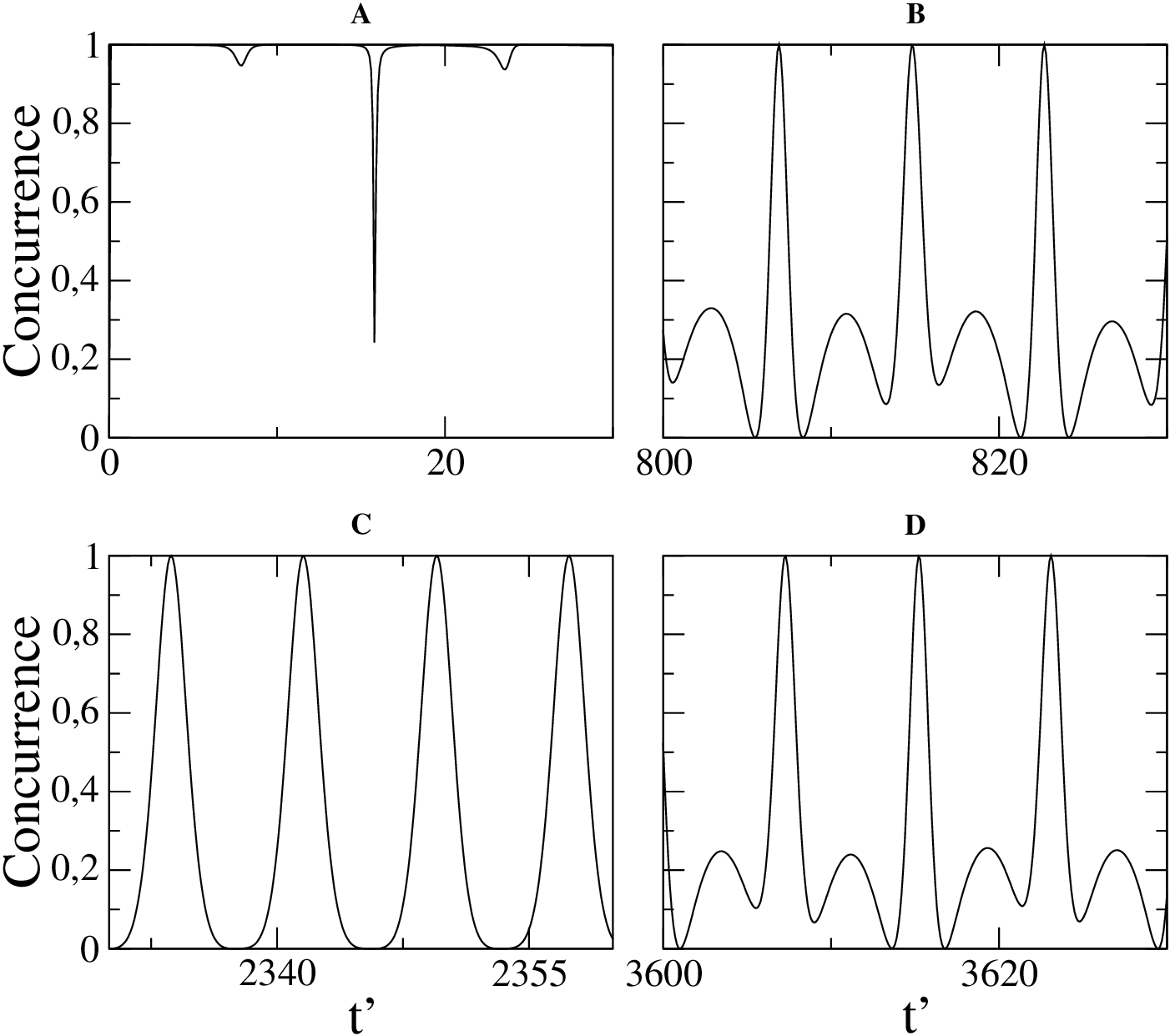}
\caption{The result of entanglement swapping for different values of the parameters $\omega_2 = \Omega_2 = 0.8$, $\omega_1=\Omega_1 = 1,~g_k=0.2$, The initial state is $ \vert\psi(0)\rangle^{tot}=\vert \downarrow 0 \rangle^1\otimes\vert \downarrow 0\rangle^2$.}
\label{f5}
\end{figure}

We have also studied the entanglement for the subsystems  having different frequencies in longer time range $t'$. We have found an interesting cyclic behaviour shown in Figs.\ref{f4} and \ref{f5}.
The difference in $QC$s frequencies results in change of the character of the photons interference pattern that swaps onto the qubits entanglement. 

For the initial condition (\ref{e0g1}) the oscillations of concurrence are at first regular (Fig.\ref{f4} A), changing to a less regular (Fig.\ref{f4} B), reaching strong qubit-qubit entanglement regime (Fig.\ref{f4} C).  With further increasing $t'$ the concurrence exhibits peaked structure again (Fig.\ref{f4} D) and returns back to the regular oscillations of the Fig.\ref{f4} A. Surprisingly, starting from the initial state (\ref{e0e0}) the behaviour is analogous but shifted in time. It is shown in Fig.\ref{f5}.  

\section{Conclusions}
We have discussed a physical interface between quantum microwaves and solid-state system based on interactions between field modes and non-superconducting flux qubits. We have examined the problem of transfer of entanglement between such systems for various realistic configurations. In particular, we have studied the influence of different initial states and the qubits frequencies on the swapping procedure. We have found that the multi-excitation initial states influence destructively on swapping of entanglement, however the qubits still get entangled. This influence is less pronounced if the $QC$s start from the same initial states. In this case there exists, at least in theory, a state for which the presence on many excitations does not change the single excitation results.  \\
For the system with a single excitation when both $QC$ subsystems are not in resonance the concurrence exhibits interesting cyclic behaviour as a function of the time $t'$ at which the BSM has been performed.

It follows from our model calculations that to get the coherent coupling of remotely located qubits it is much more favourable if the $QC$ subsystems start from the same initial states. This process is then robust against the change of the number of excitations and qubit frequencies.

In this paper we have concentrated on the analysis of coherent coupling of qubits neglecting the influence of dissipation. However as was shown in \cite{zipper,kurpas} decoherence decreases the amplitude of entanglement but does not kill the phenomenon for experimentally feasible parameters.
The process of entanglement through a joint detection of photons turns out to be a reliable mechanism for the engineering entanglement between two solid state qubits.   

\begin{acknowledgement}
We thank M. \.Zukowski for valuable comments and discussion. Work supported by the Scientific Network LEPPI Nr 106/E-345/BWSN-0166/2008, by the Polish Ministry of Science and Higher Education under the grant N 202 131 32/3786 and the scholarship from the UPGOW project co-financed by the European Social Fund.
\end{acknowledgement}

\end{document}